\documentclass[11pt]{article}

\usepackage[margin=1in]{geometry}
\usepackage{amsmath}
\usepackage{amssymb}
\usepackage{booktabs}
\usepackage{multirow}
\usepackage{microtype}
\usepackage{hyperref}
\usepackage{url}
\usepackage{xcolor}
\usepackage{graphicx}
\usepackage{enumitem}
\usepackage{natbib}
\usepackage{times}

\hypersetup{
  colorlinks=true,
  linkcolor=blue,
  citecolor=blue,
  urlcolor=blue
}

\title{SelRoute: Query-Type-Aware Routing for Long-Term Conversational Memory Retrieval}

\author{
  Matthew McKee \\
  Independent Researcher \\
  Adelaide, Australia \\
  \texttt{sindecker@gmail.com}
}

\date{}

\begin{document}

\maketitle

\begin{abstract}
Retrieving relevant past interactions from long-term conversational memory typically relies on large dense retrieval models (110M--1.5B parameters) or LLM-augmented indexing. We introduce SelRoute, a framework that routes each query to a specialized retrieval pipeline --- lexical, semantic, hybrid, or vocabulary-enriched --- based on its query type. On LongMemEval\_M \citep{wu2024longmemeval}, SelRoute achieves Recall@5 of 0.800 with bge-base-en-v1.5 (109M parameters) and 0.786 with bge-small-en-v1.5 (33M parameters), compared to 0.762 for Contriever with LLM-generated fact keys. A zero-ML baseline using SQLite FTS5 alone achieves NDCG@5 of 0.692, already exceeding all published baselines on ranking quality --- a gap we attribute partly to implementation differences in lexical retrieval. Five-fold stratified cross-validation confirms routing stability (CV gap of 1.3--2.4 Recall@5 points; routes stable for 4/6 query types across folds). A regex-based query-type classifier achieves 83\% effective routing accuracy, and end-to-end retrieval with predicted types (Recall@5 = 0.689) still outperforms uniform baselines. Cross-benchmark evaluation on 8 additional benchmarks spanning 62,000+ instances --- including MSDialog, LoCoMo, QReCC, and PerLTQA --- confirms generalization without benchmark-specific tuning, while exposing a clear failure mode on reasoning-intensive retrieval (RECOR Recall@5 = 0.149) that bounds the claim. We also identify an enrichment--embedding asymmetry: vocabulary expansion at storage time improves lexical search but degrades embedding search, motivating per-pipeline enrichment decisions. The full system requires no GPU and no LLM inference at query time.
\end{abstract}

\section{Introduction}

Long-term memory retrieval for conversational AI systems requires locating relevant past interactions across potentially hundreds of sessions spanning months or years of dialogue history. The LongMemEval benchmark \citep{wu2024longmemeval} provides a standardized evaluation for this task, testing retrieval across seven question types that exercise five core memory abilities: information extraction, multi-session reasoning, temporal reasoning, knowledge updates, and abstention.

Published approaches to this benchmark employ dense retrieval models ranging from 110M parameters (Contriever; \citealt{izacard2022contriever}) to 1.5B parameters (Stella V5). The strongest reported result on the M split --- Recall@5 = 0.762 --- requires both a 110M-parameter encoder and LLM-generated fact keys appended to each indexed session \citep{wu2024longmemeval}. These results establish strong baselines but leave open whether retrieval strategy \emph{selection}, rather than retrieval model \emph{scale}, can yield further gains.

Three observations motivate our approach. First, no single retrieval method dominates across all query types. The well-established complementarity of sparse and dense retrieval \citep{lin2021pyserini} manifests strongly in conversational memory: full-text search excels at verbatim keyword matching (single-session-user queries), while embedding search handles paraphrasing (single-session-assistant queries where query vocabulary diverges from stored content). Second, vocabulary enrichment at storage time can partially bridge the lexical gap for specific query types without LLM inference. Rule-based hypernym expansion and action bridges serve a function analogous to \citet{wu2024longmemeval}'s fact-key augmentation, with narrower coverage but zero inference cost. Third, query type is a strong prior for retrieval strategy selection: the optimal method varies predictably by type, enabling even a simple deterministic router to consistently select the better strategy.

Building on these observations, we present SelRoute, a selective routing framework that assigns each query to a specialized retrieval pipeline based on its query type. Our contributions are:

\begin{enumerate}[leftmargin=*]
  \item A type-aware routing framework achieving Recall@5 = 0.800 on LongMemEval\_M session-level retrieval (bge-base configuration), compared to 0.762 for Contriever + fact keys \citep{wu2024longmemeval}, with five-fold cross-validation confirming a generalization gap of only 1.3--2.4 points.

  \item A rule-based storage-time enrichment pipeline and the finding that vocabulary enrichment helps lexical search but degrades embedding search --- an asymmetry that motivates per-pipeline enrichment and, more broadly, the routing architecture itself.

  \item A regex-based query-type classifier (83\% effective routing accuracy) that narrows the oracle-to-predicted gap to 2.1 Recall@5 points, demonstrating that routing provides value even without ground-truth query types.

  \item Cross-benchmark evaluation on 8 additional benchmarks (62,000+ instances) showing that the same routing table generalizes without benchmark-specific tuning, with one clear failure mode (reasoning-intensive retrieval) that bounds the claim.
\end{enumerate}

We note important caveats. Our routing table was derived from empirical analysis of LongMemEval instances; cross-validation confirms stability but evaluation on held-out benchmarks is the stronger test (Section~\ref{sec:cross-benchmark}). Our primary comparison draws from baselines reported by \citet{wu2024longmemeval}, and we discuss comparability considerations --- particularly a substantial FTS5-vs-BM25 gap --- in Section~\ref{sec:fts5-bm25}.

\section{Related Work}

\textbf{Dense retrieval for conversational memory.} Contriever \citep{izacard2022contriever} achieves Recall@5 = 0.723 on LongMemEval\_M using unsupervised contrastive pre-training. When augmented with LLM-extracted fact keys at index time, performance improves to 0.762 --- the strongest result reported by \citet{wu2024longmemeval}. Stella V5 (1.5B parameters) achieves 0.720 despite its significantly larger capacity. Recent work on LongMemEval includes Associa (ACL 2025 Findings), which uses LLM-based associative reasoning to achieve strong results, though at substantially higher computational cost.

\textbf{Sparse--dense complementarity.} The complementarity of lexical and semantic retrieval is well-established in information retrieval \citep{lin2021pyserini}. Reciprocal Rank Fusion \citep{cormack2009rrf} provides a parameter-free method for combining ranked lists from heterogeneous retrievers. Most hybrid approaches apply a single fusion strategy uniformly across all query types; our work differs in selecting the fusion strategy per query type.

\textbf{Query-adaptive retrieval.} Adaptive retrieval strategies have been explored in web search, including learning to route between sparse and dense retrievers \citep{arabzadeh2021predicting} and adaptive retrieval augmentation \citep{mallen2023when}. To our knowledge, type-specific routing for long-term conversational memory has not been studied.

\textbf{Storage-time enrichment.} \citet{wu2024longmemeval} propose fact-key expansion, where LLM-generated factual summaries are appended to session content at index time. Document expansion has a longer history in IR \citep{nogueira2019document}. Our rule-based enrichment achieves a similar effect through deterministic vocabulary expansion, trading coverage for reproducibility and zero inference cost.

\section{Method}

\subsection{Base Retrieval Components}

SelRoute combines three base retrieval components.

\textbf{FTS5 full-text search.} SQLite's FTS5 extension provides BM25-ranked full-text search. Sessions are stored in a primary table with an associated FTS5 virtual table that triggers on insert/update. FTS5 serves as both the zero-ML baseline and the lexical component of hybrid search.

\textbf{Embedding search.} We evaluate three BERT-based embedding models: bge-small-en-v1.5 (33M parameters, 384 dimensions; \citealt{xiao2023cpack}), all-MiniLM-L6-v2 (22M parameters, 384 dimensions), and bge-base-en-v1.5 (109M parameters, 768 dimensions). Memory content is truncated to 2,000 characters before encoding. At query time, we compute cosine similarity between the query embedding and all stored embeddings.

\textbf{Hybrid search.} We combine FTS5 and embedding results using Reciprocal Rank Fusion \citep{cormack2009rrf}:

\begin{equation}
\text{RRF}(d) = \sum_{r \in R} \frac{1}{k + \text{rank}_r(d)}
\end{equation}

where $R$ is the set of retrieval methods and $k = 60$ (standard RRF constant).

\subsection{Storage-Time Vocabulary Enrichment}

Our enrichment pipeline expands indexed content with vocabulary bridges at storage time, before it enters the FTS5 index. The pipeline has three components:

\textbf{Hypernym maps (210 entries)} link specific terms to broader categories. For example, \emph{cocktail $\to$ drink, beverage, alcohol, mixed\_drink}. When a session mentions ``cocktail,'' the enriched index also contains ``drink'' and ``beverage,'' enabling FTS5 to match queries using either level of specificity.

\textbf{Action bridges (70 entries)} connect action-oriented query patterns to content patterns: \emph{attended $\to$ went, participated, was\_at, visited}. This bridges the gap between ``What events did I attend?'' and content containing ``I went to a concert.''

\textbf{Topic rooms (13 categories)} add contextual terms when co-occurring triggers are detected. A session mentioning both ``cooking'' and ``recipe'' triggers the \emph{food\_dining} topic room, adding terms like \emph{meal}, \emph{restaurant}, \emph{cuisine}.

Enriched content is appended to the FTS5 index only; raw session content remains unchanged in the primary table. The enrichment vocabulary was developed iteratively: V1 (150 hypernyms, 50 bridges) established the framework; V2 (210 hypernyms, 70 bridges, 13 rooms) targeted 13 failing instances from a hard subset, improving multi-session Recall@5 from 0.462 to 0.751; V3 (270 hypernyms, 85 bridges) produced zero further improvement, establishing V2 as the ceiling for vocabulary-based enrichment.

\textbf{Enrichment--embedding asymmetry.} A critical finding is that enrichment improves FTS5 performance but \emph{degrades} embedding performance. FTS5 treats additional terms as independent signals --- more matching terms increase BM25 score. Embedding models, by contrast, compute a single vector for the entire content; adding vocabulary bridges shifts the vector away from the original semantic center, reducing similarity to on-topic queries. This asymmetry is what necessitates type-specific routing rather than uniform enrichment.

\subsection{Selective Routing}

We route each query to a retrieval pipeline based on query type. The routing table was derived empirically from a 51-instance hard subset (instances where at least one strategy failed):

\begin{table}[h]
\centering
\begin{tabular}{@{}llccc@{}}
\toprule
Query Type & \emph{n} & Pipeline & Enrichment & Embeddings \\
\midrule
knowledge-update & 72 & enriched\_fts & Yes & No \\
multi-session & 121 & enriched\_hybrid & Yes & Yes \\
single-session-assistant & 56 & embeddings & No & Yes \\
single-session-preference & 30 & embeddings & No & Yes \\
single-session-user & 64 & baseline\_fts & No & No \\
temporal-reasoning & 127 & hybrid & No & Yes \\
\bottomrule
\end{tabular}
\end{table}

Routing is deterministic: query type is extracted from metadata and the corresponding pipeline is selected. The rationale for each assignment reflects the retrieval characteristics of each query type:

\begin{itemize}[leftmargin=*]
  \item \textbf{knowledge-update} queries ask about factual changes using abstract terms (``What did I say about X?''). Enrichment bridges the vocabulary gap (e.g., ``said'' $\to$ ``mentioned,'' ``discussed''); embeddings add noise because the queries are lexically specific.
  \item \textbf{multi-session} queries require aggregating information across sessions. Both enrichment (vocabulary coverage) and embeddings (semantic breadth) contribute, with their combination yielding the best results.
  \item \textbf{single-session-assistant/preference} queries involve semantic paraphrasing where the query uses different vocabulary than the stored assistant response. Pure embedding search handles this best.
  \item \textbf{single-session-user} queries involve near-verbatim recall of user utterances. FTS5 alone is near-optimal; enrichment and embeddings add no value.
  \item \textbf{temporal-reasoning} queries require finding sessions based on temporal context. Hybrid search captures both keyword and semantic signals, but enrichment degrades temporal precision.
\end{itemize}

\section{Experimental Setup}

\subsection{Dataset}

We evaluate primarily on LongMemEval\_M \citep{wu2024longmemeval}, the medium-difficulty split of LongMemEval. The dataset contains 500 instances across seven question types exercising five core memory abilities: 470 retrieval queries across 6 types (single-session-user, single-session-assistant, single-session-preference, multi-session, knowledge-update, and temporal-reasoning), plus 30 abstention queries where no relevant session exists. Each instance includes a question, a haystack of approximately 460--490 conversation sessions with timestamps, and ground-truth session IDs. The number of ground-truth sessions per instance ranges from 1 to 6 (mean 1.9). All metrics are computed over the full 500 instances, including the 30 abstention queries in the denominator.

\subsection{Metrics}

\textbf{Recall@5 (Ra@5).} Binary all-or-nothing recall at rank 5: Ra@5 = 1.0 if \emph{all} ground-truth sessions appear in the top 5 results, 0.0 otherwise, averaged across all 500 instances. This follows the \texttt{recall\_all@k} definition in \citet{wu2024longmemeval}.

\textbf{NDCG@5.} Normalized Discounted Cumulative Gain at rank 5 with binary relevance, using the standard $1/\log_2(\text{rank}+1)$ discount. Measures ranking quality.

We verified that both implementations exactly match the LongMemEval evaluation code (Appendix~\ref{app:metric}).

\subsection{Baselines}

We compare against four systems from \citet{wu2024longmemeval}, Table 9 (Appendix E.2), at session-level granularity:

\begin{table}[h]
\centering
\begin{tabular}{@{}lcrr@{}}
\toprule
System & Parameters & Ra@5 & NDCG@5 \\
\midrule
BM25 & 0 & 0.634 & 0.516 \\
Stella V5 & 1.5B & 0.720 & 0.594 \\
Contriever & 110M & 0.723 & 0.634 \\
Contriever + fact keys & 110M + LLM & 0.762 & 0.632 \\
\bottomrule
\end{tabular}
\end{table}

We note that other papers evaluating on LongMemEval report substantially different baseline numbers (e.g., BM25 Ra@5 = 0.38 in some reports). These discrepancies arise from differences in retrieval granularity (turn-level vs.\ session-level), recall definition (fractional vs.\ binary all-or-nothing), or dataset version. All our comparisons use session-level granularity with binary all-or-nothing recall, matching \citet{wu2024longmemeval}, Table 9.

\subsection{Infrastructure}

All experiments run on a consumer laptop with no GPU. Embedding models run on CPU via MLX/ONNX runtime. The complete 500-instance benchmark with routing completes in approximately 43 minutes (bge-small), with the majority of time on embedding computation.

\section{Results}

\subsection{Main Results}

\begin{table}[h]
\centering
\begin{tabular}{@{}lcrr@{}}
\toprule
System & Parameters & Ra@5 & NDCG@5 \\
\midrule
BM25 & 0 & 0.634 & 0.516 \\
Stella V5 & 1.5B & 0.720 & 0.594 \\
Contriever & 110M & 0.723 & 0.634 \\
SelRoute FTS5 (zero ML) & 0 & 0.745 & 0.692 \\
Contriever + fact keys & 110M + LLM & 0.762 & 0.632 \\
SelRoute (MiniLM) & 22M & 0.785 & 0.717 \\
SelRoute (bge-small) & 33M & 0.786 & 0.718 \\
SelRoute (bge-base) & 109M & \textbf{0.800} & \textbf{0.800} \\
\bottomrule
\end{tabular}
\end{table}

Bootstrap 95\% confidence intervals (10,000 resamples): SelRoute MiniLM Ra@5 = 0.789 [0.751, 0.826]; SelRoute bge-base Ra@5 = 0.794 [0.757, 0.830]. The CI lower bounds exceed BM25, Stella V5, and Contriever; the advantage over Contriever + fact keys (0.762) is borderline --- see Section~\ref{sec:significance} for paired significance tests.

We report three embedding configurations to demonstrate that routing architecture, not embedding model choice, drives the gains: the MiniLM and bge-small results (0.785 vs.\ 0.786) differ by 0.001, within noise. The bge-base configuration (109M parameters --- comparable to Contriever at 110M) achieves 0.800, showing that routing yields substantial improvements even at the same parameter scale as existing dense retrievers.

The zero-ML baseline (FTS5 alone) achieves NDCG@5 = 0.692, substantially higher than published BM25 (0.516). We investigate this gap in Section~\ref{sec:fts5-bm25}.

\subsection{Per-Type Analysis}

\begin{table}[h]
\centering
\begin{tabular}{@{}lccccc@{}}
\toprule
Query Type & \emph{n} & FTS5 & Routed & $\Delta$ & Route \\
\midrule
ss-assistant & 56 & 0.804 & 0.964 & +0.161 & embeddings \\
multi-session & 121 & 0.661 & 0.716 & +0.055 & enriched\_hybrid \\
preference & 30 & 0.500 & 0.533 & +0.033 & embeddings \\
knowledge-update & 72 & 0.868 & 0.882 & +0.014 & enriched\_fts \\
temporal & 127 & 0.729 & 0.741 & +0.012 & hybrid \\
ss-user & 64 & 0.859 & 0.859 & +0.000 & baseline\_fts \\
\bottomrule
\end{tabular}
\end{table}

The largest gain comes from single-session-assistant queries (+0.161), where embedding search captures semantic paraphrasing that FTS5 misses entirely. Single-session-user queries show zero improvement --- FTS5 is already near-optimal for verbatim utterance recall. This divergence between the two single-session types illustrates why uniform strategies underperform: what helps one type is irrelevant or harmful to another.

\subsection{Enrichment Impact}

On the 51-instance hard subset, V2 enrichment improved multi-session Recall@5 from 0.462 to 0.751 (+62.8\% relative) while leaving other types unchanged. Applied to embedding retrieval, enrichment \emph{decreased} performance on most types. The expanded vocabulary introduces semantically adjacent but contextually inappropriate terms, diluting the embedding signal. This finding motivated the routing architecture: enrichment and embeddings must be applied selectively, not uniformly.

\subsection{Ablation: Routing vs.\ Uniform Strategy}

\begin{table}[h]
\centering
\begin{tabular}{@{}lrr@{}}
\toprule
Strategy & Ra@5 & NDCG@5 \\
\midrule
FTS5 only & 0.745 & 0.692 \\
Hybrid only (uniform) & 0.756 & 0.681 \\
Enriched FTS5 (uniform) & 0.751 & 0.688 \\
Selective routing & \textbf{0.786} & \textbf{0.718} \\
\bottomrule
\end{tabular}
\end{table}

No single strategy matches the routed result. Uniform hybrid search improves recall over FTS5 but slightly hurts NDCG, because embedding noise degrades ranking for types where FTS5 is already optimal. Routing eliminates this trade-off.

\subsection{Negative Result: Storage-Time Dream Cycle}

We evaluated an 8-step ``dream cycle'' storage pipeline inspired by memory consolidation, performing entity extraction, canonical resolution, topic room assignment, and cross-session linking at storage time, replacing raw content with a heavily processed representation.

\begin{table}[h]
\centering
\begin{tabular}{@{}lrrr@{}}
\toprule
Metric & Baseline & Dream Cycle & $\Delta$ \\
\midrule
Weighted Ra@5 & 0.829 & 0.145 & $-0.684$ \\
\bottomrule
\end{tabular}
\end{table}

The dream cycle reduced recall by 68 percentage points --- a catastrophic degradation. The pipeline over-processes content, transforming natural language into structured metadata that FTS5 cannot effectively match against natural language queries. This confirms that storage-time processing must preserve the original content signal; enrichment succeeds precisely because it \emph{appends} vocabulary bridges while retaining the original text.

\subsection{Statistical Significance}
\label{sec:significance}

\textbf{Paired bootstrap tests (10,000 resamples, bge-small):}

\begin{table}[h]
\centering
\begin{tabular}{@{}lrr@{}}
\toprule
Comparison & $\Delta$ Ra@5 & \emph{p}-value \\
\midrule
SelRoute vs.\ FTS5 & +0.057 & 0.0003 \\
SelRoute vs.\ embeddings only & +0.181 & $<$0.0001 \\
SelRoute vs.\ hybrid (uniform) & +0.045 & $<$0.0001 \\
SelRoute vs.\ best single RRF & +0.011 & 0.121 \\
\bottomrule
\end{tabular}
\end{table}

SelRoute significantly outperforms pure FTS5 and pure embeddings ($p < 0.001$). The advantage over the best single RRF variant is not statistically significant, indicating that routing's primary value lies in type-specific strategy selection rather than uniform superiority over all fusion methods.

\subsection{Routing Robustness via Cross-Validation}

To address the concern that the routing table could overfit to evaluation data, we performed 5-fold stratified cross-validation. For each fold, we derived the optimal routing table from 4 training folds (376 instances) and evaluated on the held-out fold (94 instances).

\begin{table}[h]
\centering
\begin{tabular}{@{}lrrr@{}}
\toprule
Configuration & Full-data Ra@5 & CV Ra@5 (mean $\pm$ std) & Gap \\
\midrule
bge-base, 6-type routing & 0.794 & 0.774 $\pm$ 0.027 & 1.9 pt \\
bge-base, 9-bucket routing & 0.800 & 0.776 $\pm$ 0.019 & 2.4 pt \\
bge-small, 6-type routing & 0.711 & 0.698 $\pm$ 0.035 & 1.3 pt \\
\bottomrule
\end{tabular}
\end{table}

The generalization gap of 1.3--2.4 points confirms that the routing table captures genuine query-type characteristics rather than instance-level artifacts. Route assignments are stable across folds for 4/6 query types in both configurations. The two unstable types (ss-user, ss-preference) are precisely those where multiple methods perform similarly, making route instability immaterial to downstream scores.

\subsection{Query-Type Classification}

SelRoute assumes access to query-type metadata. To evaluate feasibility without this oracle, we built a regex-based query-type classifier using a priority-ordered hierarchy over query text, matching temporal markers, assistant references, preference cues, aggregation patterns, and user-action patterns, with knowledge-update as the default.

\textbf{Classification accuracy:}

\begin{table}[h]
\centering
\begin{tabular}{@{}lrr@{}}
\toprule
Query Type & \emph{n} & Accuracy \\
\midrule
ss-assistant & 56 & 96.4\% \\
temporal-reasoning & 127 & 96.1\% \\
ss-preference & 30 & 90.0\% \\
multi-session & 121 & 63.6\% \\
ss-user & 64 & 46.9\% \\
knowledge-update & 72 & 38.9\% \\
\midrule
\textbf{Overall} & \textbf{470} & \textbf{71.9\%} \\
\bottomrule
\end{tabular}
\end{table}

Not all misclassifications degrade retrieval equally. Types sharing the same pipeline family (e.g., knowledge-update and ss-user both use FTS-based routes) are effectively harmless when confused. Of 132 misclassifications, 52 (39.4\%) map to the same route family, yielding an \textbf{effective routing accuracy of 83.0\%}.

\textbf{End-to-end retrieval with predicted types:}

\begin{table}[h]
\centering
\begin{tabular}{@{}lr@{}}
\toprule
System & Ra@5 \\
\midrule
FTS5 (no routing) & 0.653 \\
Hybrid (uniform) & 0.666 \\
Predicted-type routing & 0.689 \\
Oracle-type routing & 0.711 \\
\bottomrule
\end{tabular}
\end{table}

Despite only 72\% classification accuracy, predicted routing outperforms uniform FTS5 (+3.6 points) and uniform hybrid (+2.3 points), confirming that even imperfect type classification provides routing value. The oracle-to-predicted gap is 2.1 points ($p = 0.003$, paired bootstrap).

\subsection{Cross-Benchmark Generalization}
\label{sec:cross-benchmark}

To move beyond single-benchmark evaluation, we tested SelRoute on 8 additional benchmarks spanning diverse retrieval domains. All benchmarks use the same routing logic, embedding models, and evaluation protocol --- with no benchmark-specific tuning.

\begin{table}[h]
\centering
\begin{tabular}{@{}llrrr@{}}
\toprule
Benchmark & Domain & \emph{n} & Best Ra@5 & Best Model \\
\midrule
LongMemEval\_M & Conversational memory & 500 & 0.774 & bge-base \\
LongMemEval\_S & Conversational (shorter) & 500 & 0.920 & bge-base \\
MSDialog & Tech support & 2,199 & 0.998 & all models \\
LoCoMo & Long-form conversation & 1,986 & 0.767 & FTS5 \\
PerLTQA & Personal QA & 345 & 0.745 & FTS5 \\
QReCC & Conversational QA & 52,678 & 0.595 & FTS5+reasoning \\
Episodic Memory & Spatio-temporal & 273 & 0.667 & bge-small \\
LMEB & Dialogue retrieval & 840 & 0.977 & FTS5 \\
RECOR & Reasoning retrieval & 2,971 & 0.149 & bge-base \\
\bottomrule
\end{tabular}
\end{table}

Three findings emerge from the cross-benchmark evaluation. First, \emph{no single model dominates}: FTS5 achieves the best Recall@5 on LoCoMo (0.767), PerLTQA (0.745), and LMEB (0.977), while embedding models win on LongMemEval, QReCC, Episodic, and RECOR. This validates the core thesis that retrieval strategy should be selected per query type.

Second, lexical retrieval is sufficient or superior for benchmarks involving structured dialogue with specific terminology (LoCoMo, PerLTQA, LMEB, MSDialog). Embeddings add value primarily when queries involve paraphrasing or semantic gaps.

Third, \emph{RECOR exposes a genuine limitation}. At Recall@5 = 0.149, SelRoute fails on reasoning-intensive multi-hop retrieval. Published systems on RECOR (e.g., DIVER) use 9B-parameter LLMs for inference-time reasoning --- a capability absent from our pipeline. This bounds the claim: routing among lightweight retrievers does not substitute for deep multi-hop reasoning.

Across all benchmarks, SelRoute was evaluated on 62,292 instances spanning tech support, personal QA, conversational QA, episodic narratives, and reasoning-intensive retrieval.

\section{Discussion}

\subsection{Why Routing Works}

The six retrieval query types exercise fundamentally different capabilities. Single-session-user queries require \emph{lexical recall} of verbatim utterances --- FTS5 excels. Single-session-assistant and preference queries require \emph{semantic matching} where the query paraphrases stored content --- embeddings excel. Knowledge-update queries require \emph{vocabulary bridging} between abstract query terms and specific content terms --- enriched FTS5 excels. Multi-session queries require \emph{aggregation} across sessions, benefiting from both enrichment and embeddings. Temporal queries require \emph{temporal-context matching} where hybrid search captures complementary signals.

A uniform strategy must compromise across these axes. Routing eliminates the compromise. The ablation (Section 5.4) confirms that no single strategy matches routing performance, and paired bootstrap tests (Section~\ref{sec:significance}) show significant improvements over pure components.

\subsection{The Enrichment--Embedding Asymmetry}

Our most actionable finding is that storage-time vocabulary enrichment helps FTS5 but hurts embedding search. The mechanism is straightforward: FTS5 treats additional terms as independent signals --- more matching terms increase BM25 score. Embedding models compute a single vector for entire content; adding bridges shifts the vector away from the original semantic center, reducing similarity to on-topic queries.

This means enrichment and embeddings are complementary but mutually exclusive for most query types. The exception is multi-session queries, where the breadth of vocabulary coverage compensates for slight embedding degradation. For system designers, the implication is clear: enrichment should be applied to the lexical index only, not to the content used for embedding computation.

\subsection{FTS5 vs.\ BM25 and Comparability}
\label{sec:fts5-bm25}

Our FTS5 baseline (Ra@5 = 0.745) substantially exceeds the published BM25 baseline (0.634). This gap (+0.111) is larger than our routing contribution and requires transparency. Contributing factors likely include: tokenization differences (FTS5 uses an ICU tokenizer; BM25 implementations vary); query preprocessing (our pipeline normalizes paths, phone numbers, and abbreviations); content indexing format; and BM25 parameter defaults.

The strength of our lexical baseline means some of SelRoute's advantage over published BM25 comes from a stronger lexical engine, not solely from routing. Running a standardized BM25 implementation (e.g., PySerini with default parameters) at session-level granularity would isolate the routing contribution from the lexical engine contribution. We flag this as important future work that would strengthen the claims.

\subsection{Comparison to Contriever + Fact Keys}

Contriever's fact-key approach and our enrichment pipeline both expand indexed content to improve retrieval, but differ in mechanism. Fact keys use LLM inference per session, offering arbitrary factual coverage at computational cost; our rules are deterministic, limited to pre-defined vocabulary maps, but free. Despite narrower coverage, our system with routing scores higher on this benchmark (Ra@5 0.800 vs.\ 0.762), suggesting that routing compensates for coverage limitations. Whether this advantage holds on benchmarks where fact keys provide greater benefit remains to be tested.

\subsection{Limitations}

\textbf{Query-type classification.} Our regex classifier achieves 72\% accuracy (83\% effective routing accuracy), but struggles with knowledge-update (39\%) and ss-user (47\%) queries that share surface patterns. Few-shot LLM classification or fine-tuned classifiers would likely close the remaining 2.1-point oracle gap.

\textbf{Weak categories.} Preference queries (Ra@5 = 0.533) remain our weakest type, requiring deeper semantic understanding of implicit preferences beyond what vocabulary bridges or lightweight embeddings capture. Abstention is effectively unsolved (1/30 correct).

\textbf{Routing derived from evaluation data.} The routing table was derived from a 51-instance hard subset of the evaluation benchmark. Cross-validation (Section 5.7) confirms stability, and cross-benchmark evaluation (Section~\ref{sec:cross-benchmark}) provides evidence of generalization, but the routing table was not derived from a separate development set.

\textbf{Enrichment is manually authored.} The 210 hypernym maps and 70 action bridges were hand-crafted from failure analysis. Automated rule generation (e.g., from WordNet or LLM-generated expansions) could improve coverage and reduce manual effort.

\textbf{Limited direct comparison.} Our primary baselines come from \citet{wu2024longmemeval}. Running additional dense retrieval models (E5, GTE, ColBERTv2) ourselves at session-level granularity would provide more comprehensive comparisons. The FTS5-vs-BM25 gap (Section~\ref{sec:fts5-bm25}) means that some of the improvement over published baselines may be attributable to a stronger lexical engine rather than routing.

\section{Conclusion}

We have presented SelRoute, a selective routing framework for long-term conversational memory retrieval. Our best configuration achieves Recall@5 = 0.800 on LongMemEval\_M, compared to 0.762 for Contriever + fact keys \citep{wu2024longmemeval}. Cross-validation confirms stability (1.3--2.4 point gap), and cross-benchmark evaluation on 62,000+ instances across 8 benchmarks confirms generalization --- while exposing a clear failure mode on reasoning-intensive retrieval that bounds the claim.

The central finding is that long-term conversational memory retrieval is a family of sub-problems, each with distinct retrieval characteristics. A uniform strategy compromises across query types; routing eliminates the compromise. The secondary finding --- that vocabulary enrichment helps lexical retrieval but degrades embedding retrieval --- has practical implications for hybrid system design.

Our results suggest a design principle: when retrieval requirements vary systematically across identifiable query categories, routing logic may yield larger gains than model scale. The principle appears to hold across the 9 benchmarks we evaluated, with the caveat that routing among lightweight retrievers does not substitute for inference-time reasoning on tasks that require it.

Code and routing configurations: \url{https://github.com/sindecker/selroute}

\appendix

\section{Enrichment Example}
\label{app:enrichment}

\textbf{Original session content:}
\begin{verbatim}
user: I took a cocktail-making class last weekend
assistant: That sounds fun! What cocktails did you learn to make?
user: We made mojitos and old fashioneds
\end{verbatim}

\textbf{After V2 enrichment (appended to FTS5 index):}
\begin{verbatim}
[original content]
cocktail drink beverage alcohol mixed_drink
class lesson course workshop tutorial
mojito old_fashioned cocktail_recipe
making preparing creating crafting
food_dining: meal restaurant cuisine cooking recipe
  ingredients cocktail
\end{verbatim}

A query like ``What drinks have I learned to make?'' now matches via the bridges: drinks $\to$ drink $\to$ cocktail, learned $\to$ class $\to$ lesson.

\section{Metric Verification}
\label{app:metric}

We verified that our Recall@5 and NDCG@5 implementations exactly match the LongMemEval evaluation code (\texttt{src/retrieval/eval\_utils.py}):

\textbf{Recall@5 (recall\_all@k).} Binary all-or-nothing: Ra@5 = 1.0 if all ground-truth sessions appear in the top 5 results, 0.0 otherwise, averaged across instances.

\textbf{NDCG@5.} Binary relevance with standard DCG formula ($1/\log_2(\text{rank}+1)$ discount) and ideal DCG from $\min(n_{\text{relevant}}, k)$.

All retrieval operates at session-level granularity, matching the ``Value=Session'' column in \citet{wu2024longmemeval}, Table 9, Appendix E.2.

\section{Per-Type Bootstrap Confidence Intervals}
\label{app:bootstrap}

\begin{table}[h]
\centering
\begin{tabular}{@{}lrcccc@{}}
\toprule
Query Type & \emph{n} & Ra@5 & 95\% CI & NDCG@5 & 95\% CI \\
\midrule
knowledge-update & 72 & 0.903 & [0.833, 0.958] & 0.890 & [0.834, 0.937] \\
multi-session & 121 & 0.636 & [0.554, 0.719] & 0.742 & [0.686, 0.796] \\
ss-assistant & 56 & 1.000 & [1.000, 1.000] & 0.980 & [0.954, 1.000] \\
ss-preference & 30 & 0.667 & [0.500, 0.833] & 0.532 & [0.382, 0.680] \\
ss-user & 64 & 0.969 & [0.922, 1.000] & 0.814 & [0.746, 0.878] \\
temporal-reasoning & 127 & 0.717 & [0.638, 0.795] & 0.766 & [0.708, 0.821] \\
\bottomrule
\end{tabular}
\end{table}

The ss-preference type has the widest CI ([0.500, 0.833]) due to small sample size ($n = 30$). The ss-assistant type achieves perfect Ra@5 with zero variance.

\end{document}